\documentstyle[prb,aps]{revtex}
\tighten
\begin{document}
\begin{titlepage}
\title{Crossover from the 2D Heisenberg to the 1D Quantum Spin Ladder
Regime in Underdoped High $T_c$ Cuprates}
\author{V.V.~Moshchalkov\\*[3mm]
Laboratorium voor Vaste-Stoffysica en Magnetisme,\\ Celestijnenlaan~200~D,
3001~Leuven, Belgium}
\maketitle
\thispagestyle{empty}
\begin{abstract}
The enigmatic scaling behaviour of the normal state properties of the high
$T_c$  cuprates has been explained by assuming that a crossover from the
two-dimensional Heisenberg (2D-H) to the one-dimensional spin ladder
(1D-SL) regime takes place at temperature $T \simeq T^*$.  For $T < T^*$
stripe formation results in the quantum 1D transport with the
characteristic inelastic length $L_\phi$ being fully controlled by the
magnetic correlation length $\xi_m$ of the even-chain SL, whereas for $T
> T^*$ the 2D quantum transport is realized with $L_\phi$ governed by the
2D-H correlations $L_\phi \simeq \xi_m \sim \exp(J/T)$.  Therefore,
the pseudogap found in underdoped ($p < p_{\mbox{\footnotesize opt}}$) high
$T_c$'s is the spin-gap $\Delta(p)$ in even-chain 1D-SL.
\end{abstract}
\vfill

\end{titlepage}
\newpage
During the last decade an impressive progress has been made in developing
an adequate description of the evolution of the high $T_c$'s normal state
properties with the hole doping $p$ (for recent overviews see Ref.~1--5).
For underdoped ($p < p_{\mbox{\footnotesize opt}}$) high $T_c$ cuprates a
convincing scaling behaviour has been demonstrated for transport [6--8],
NMR [9,10] and other data.  The established generalized $T - p$ phase
diagram contains one (or two) characteristic scaling temperatures $T^*(p)$
decreasing with $p$ and associated with the existence of the pseudogap at
$T < T^*(p)$.  While the existence of the $T^*(p)$ line on the $T -
X$ diagram is by now widely accepted, the origin of $T^*(p)$ and the
pseudogap has not yet been clarified.

The dependence $T^*(p)$ can be easily found from the scaling analysis of
the normal state resistivity $\rho(T,p)$ by taking $T^*$ as the temperature
below which a pronounced deviation of $\rho$ from the linear $\rho$ vs $T$
behaviour is seen.  Therefore, the key factor in revealing the origin of
$T^*$ is the identification of the mechanisms responsible for the specific
$\rho$ vs $T$ variation at both $T > T^*$ (linear $\rho(T)$) and $T < T^*$
(superlinear $\rho(T)$).  Generalizing the ideas formulated for high
$T_c$'s in the previous publication [11], it will be assumed here that
(i)~the
dominant scattering mechanism in the whole temperature range is of magnetic
origin; (ii)~the specific temperature dependence of resistivity, $\rho(T)$,
can be described by the inverse quantum conductivity $\sigma^{-1}$ with the
inelastic
length $L_\phi$ being fully controlled, via a strong interaction of holes
with Cu$^{2+}$ spins, by the magnetic correlation length $\xi_m$, and,
finally, (iii)~2D and 1D expressions should be used for calculating
conductivity at $T > T^*$ and $T < T^*$, respectively:
\begin{eqnarray}
\rho_{2D}^{-1}(T) &=& \left.\sigma_{2D}(T)\right|_{L_\phi =
\xi_m} = \frac{1}{b}\,\frac{e^2}{\hbar} \ln (\xi_{m2D}(T)/\ell)
\label{1}\\*[5mm]
\rho_{1D}^{-1}(T) &=& \left.\sigma_{1D}(T)\right|_{L_\phi = \xi_m} =
\frac{1}{b}\,\frac{e^2}{\hbar} \xi_{m1D}(T)
\label{2}
\end{eqnarray}
Here $\ell$ is the elastic length and $b$ is the thickness of the 2D film
or diameter of the 1D wire [12].  In Ref.~11 it has been shown that the
high temperature ($T > T^*$) transport properties of the high $T_c$
cuprates can be successfully explained in terms of the quantum 2D transport
with $L_\phi$ controlled by $\xi_m$ corresponding to the 2D Heisenberg
magnetic correlation length [13,14]:
\begin{equation}
\xi_{m2D}(T) = \frac{e}{8}\,\frac{\hbar c}{2\pi F^2}
\left(1 - \frac{1}{2}\,\frac{T}{2 \pi F^2}\right) \exp
\left(\frac{2 \pi F^2}{T}\right)
\label{3}
\end{equation}
with $2 \pi F^2 \simeq J$.  In this case the most remarkable
transport property~-- linear $\rho(T)$ behaviour~-- is simply found by
taking $\ln$ (Eq.~\ref{1}) from the exponent (Eq.~\ref{3})
\[\sigma_{2D} \sim \ln \xi_m \sim \ln\left(\exp
\frac{J}{T}\right) \sim \frac{J}{T}\]
which gives a linear $\rho$ vs $T$ dependence with the slope determined
through the exchange coupling $J$ [11].
The transition from convex (underdoped) via linear (optimally doped) to
concave (overdoped) shape of the $\rho(T)$ curves, induced by doping, as
well as the absolute $\rho(T)$ value, are also reproduced in this case
[11].

In the present paper it will be shown that the application of the 1D
scenario (Eq.~\ref{2}) below $T^*$ with $\xi_m$ determined by the
correlation length for the quantum spin ladder (SL) with an even number of
legs $n_c$ [15]
\begin{equation}
(\Delta \xi_{m1D})^{-1} = 2/\pi + A(T/\Delta) \exp (-\Delta/T)
\label{4}
\end{equation}
in combination with the 2D scenario at $T > T^*$ (Eq.~\ref{1} and
Eq.~\ref{3}) gives a convincing and consistent explanation of the whole
scaled dependence $\rho/\rho_0(T_0) = f(T/T_0)$ with $T^* \simeq 0.8
T_0$.  This proves that $T^*(p)$ corresponds to the crossover temperature
between the high temperature 2D Heisenberg and the low temperature 1D SL
regime existing, most probably, at $T_c(p)<T<T^*(p)$. This implies also the formation of the 1D stripes in the
CuO$_2$ planes.  The validity of the $\sigma_{1D} \sim \xi_{m1D}$
approach has been checked directly by applying it first to the SL cuprate
Sr$_{2.5}$Ca$_{11.5}$Cu$_{24}$O$_{41}$ where 1D stripes are just imposed by
the available crystal structure.  This approach has resulted in the fit of
a remarkable quality.  Finally, the scaling behaviour of the Knight shift in YBa$_{2}$Cu$_{3}$O$_{x}$ has also been interpreted in the framework of the 1D stripes scenario.

A rapidly growing experimental evidence, indicating the possibility of the
stripes formation [1,3,16,17], stimulates the interest to the 1D
mechanism of conductivity via charge stripes forming the lowest resistance
path.  Therefore it seems to be very important to calculate conductivity in
the 1D regime, keeping in mind an emerging reality of the stripes formation
in the 2D CuO$_2$ planes of the high $T_c$ cuprates [3,16,17].  However,
at least at this stage, it is still a bit uncomfortable to apply a pure 1D
model to materials where Cu and O orbitals form a well defined 2D system.

To overcome this psychological barrier and to verify the validity of the
proposed 1D SL model $\sigma_{1\alpha} \sim \xi_{m1D}$, we will use
it first for the description of the resistivity data obtained on the novel
even-chain spin ladder compound Sr$_{2.5}$Ca$_{11.5}$Cu$_{24}$O$_{41}$
[18].
This compound definitely contains two-leg ($n_c = 2$) Cu$_2$O$_3$ ladder
and therefore its resistivity along the ladder direction should indeed obey
Eq.~\ref{2} with $\xi_{m1D}$ given by the recent Monte Carlo
calculations [15], through the admixture of the linear temperature
dependence of $\xi_{m1D}^{-1}$ with the exponential term containing the
spin gap $\Delta$ and constant $A \simeq 1.7$ (see Eq.~\ref{4}).  The
results of the fit are shown in Fig.~1.  This fit demonstrates a remarkable
quality over the whole temperature range $T \simeq 25$--300~K, except for
the lowest temperatures where the onset of the localization effects, not
considered here, is clearly visible in experiment [18].  Moreover, the used
fitting parameters $\rho_0$, $C$ and $\Delta$ in
\begin{equation}
\rho(T) = \rho_0 + CT \exp \left(-\frac{\Delta}{T}\right) = \frac{\hbar
b^2}{e^2a} \left\{\frac{2\Delta}{\pi J_\|} + A
\frac{T}{J_\|} \exp \left(-\frac{\Delta}{T}\right)\right\}
\label{5}
\end{equation}
all show very reasonable values.  The expected residual resistence
${\displaystyle \rho_0 = \frac{\hbar
b^2}{e^2a}\,\frac{2\Delta}{\pi J_\|}}$ for $b \sim 2a \sim
7.6$~\AA, $\Delta \sim 200$~K (Fig.~1) and $J_\| \simeq 1400$~K (the normal
value for the CuO$_2$ planes) is $\rho_0 \simeq 0.5\,\cdot\,10^{-4}\
\Omega.$cm which is in a good agreement with $\rho_0 = 0.824 \times
10^{-4}\ \Omega$cm found from the fit.  The fitted gap $\Delta \simeq
213$~K (Fig.~1) is close to $\Delta = 320$~K determined for the undoped SL
SrCu$_2$O$_3$ from the inelastic neutron scattering experiments [19].  In
doped systems it is natural to expect the reduction of the spin gap.
Therefore the difference between the fitted value (213~K) and the measured
one in an undoped system (320~K) seems to be quite fair.  Finally the
calculated fitting parameter $C$ is ${\displaystyle C = \frac{\rho_0
A\pi}{2\Delta} = 0.0103}$ (in units of $10^{-4}\ \Omega$cm/K) to be
compared with $C = 0.0182$ (see Fig.~1).  Fixing the $C$ value via known
$\rho_0$ and $\Delta$ produces the fit of a comparable quality, but this
time only with the two fitting parameters $\rho_0$ and $\Delta$.

Having the validity of Eq.~\ref{5} checked on the real 1D even-chain SL
compounds, it is worth to try to use it at temperatures $T < T^*$ in $2D$
high $T_c$'s where 1D stripes might be formed.  Now it is not evident at
all that the scaled $\rho/\rho(T_0) = f(T/T_0)$ (Fig.~2a) could be also
fitted with the 1D Eq.~\ref{5}.  From the master $\rho/\rho(T_0)$ vs
$T/T_0$ curve [7] (Fig.~2a) the averaged data points (see open circles in
Fig.~2b) were used to be fitted by Eq.~\ref{5}.  The results of the fit are
given in Fig.~2b by solid lines.  First of all, at high temperatures $T >
T^* \simeq 0.8T_0$
we clearly see the anticipated linear $2D$ behaviour [11] $\rho \sim T/J$
with $J \sim T_0$.  Secondly, at $T < T^*$ the fit with Eq.~\ref{5} with
the argument $t = T/T_0$ again gives an excellent result.  It is also
clearly seen, by comparing $\rho(T)$ curves at $T < T^*$ with $\rho(T)$
curves measured for 1D SL compounds (Fig.~1), that the concave segment of
the high $T_c$ $\rho(T)$ curve at $T < T^*$ looks very much like the
$\rho(T)$ curve for the 1D even-chain SL systems.  The fitting parameter
$\Delta$, in units of $T_0$, is now $0.49\ T_0$, in agreement with the
spin-gap $\Delta \simeq 0.41\ J_\perp$ calculated for the weak-coupling
regime of the $n_c = 2$ SL [15].

Therefore, the scaling behaviour of resistivity below $T^*$
($\rho \sim \xi_{m1D}$) simply reflects the scaling curve
anticipated
for the magnetic coherence length $\xi_{m1D}$ in the $n_c = 2$ SL (see
Fig.~5 in Ref.~15).  Then the observed decrease of $T^*(p)$ (or $T_0(p)$)
[1,4,6--10] should be related to the reduction of $J_\perp$ and the spin
gap $\Delta$ in SL with doping.  Since the holes are concentrated in
stripes corresponding to SLs, then the doping level, affecting $J_\perp$
and
$J_\|$ in SL, could differ substantially and therefore one should expect
smaller $J_\perp$ and $J_\|$ than the exchange constant $J$ in the
surrounding Mott insulator phase from where the charges are expelled into
the stripes.

The formation of stripes brings into the system an intrinsic doping
inhomogeneity resulting in the reduction of dimensionality from 2D to 1D.
In resistivity measurements the most conducting paths (stripes) are
selected automatically and this makes the $\rho(T)$ data so sensitive to
the formation of stripes.  However, it is not clear to what extent stripes
dominate other properties which may integrate the response from both
stripes and the surrounding Mott insulator phase.  In this respect it is
important to apply the same approach in analyzing other scaled physical
properties.  Since in YBa$_2$Cu$_3$O$_x$ the same scaling temperature
$T_0(p)$ works equally well for resistivity and the Knight shift data [7],
the latter can also be used for fitting with the 1D SL model.

As a result, the susceptibility $\chi(T)$ (and $K(T)$) should obey the
following dependence [15,19]:
\begin{equation}
\chi(n_c,J_\perp;T) \sim T^{-1/2}\exp \: (-\Delta/T)\,.
\label{6}
\end{equation}
Fitting of the scaled $K(T)$ data with this expression (Fig.~3) gives a
good result, though requires the second fitting parameter $K_0$~-- the
Knight shift at $T \rightarrow 0$.  In this case the obtained gap
$\Delta(p) = 1.24\ T_0(p)$ (Fig.~3) is higher than $\Delta(p) = 0.49\ T_0$
found
from the analysis of the scaled resistivity at $T < T^*$ (Fig.~2b).
This difference could be caused by a contribution into $K(T)$ from atoms
inside the Mott insulator domains close to the stripes.  At the same time
such a contribution should be irrelevant for the dc transport which is
normally provided by the most conducting domains, i.e.\ stripes themselves.

In summary, the novel approach to the description of the scaling behaviour
of the high $T_c$'s normal state properties and the existence of the
pseudogap has been developed.  It is based on the assumption that dc
transport in high $T_c$'s is caused by the quantum conductivity with the
inelastic length $L\phi$ fully governed by the magnetic correlation length
$\xi_m$.  Depending upon the effective dimensionality~-- 2D ($T > T^*$)
or 1D (stripes at $T < T^*$) (Fig.~4)~-- conductivity is found from
$\sigma_{2D} \sim \ln\xi_{m2D}$ (Eq.~\ref{1}) and
$\sigma_{1D} \sim \xi_{m1D}$ (Eq.~\ref{2}), respectively.  The
validity of the $\sigma_{1D} \sim \xi_{m1D}$ approach has been
checked directly using resistivity data for real SL systems.  Then it was
successfully applied also for the description of the low temperature ($T <
T^*$) behaviour of the scaled resistivity.  In the framework of the
proposed approach the pseudogap temperature $T^*(p)$ is the crossover
between the high temperature 2D Heisenberg and the 1D quantum even-chain SL
regime.  The latter is established as temperature decreases and charge
expulsion from the Mott insulator phases [20] gets stronger and finally
leads to the formation of well defined stripes which can be modelled by the
SL with an even $n_c$.  The effective 1D case at $T < T^*$ makes all
considerations of the non-Fermi liquid behaviour [3,5,21] and possible
charge/spin separation [22,23] important for the physics of underdoped high
$T_c$'s. The weakly doped even-chain SLs have the spin-gap, show hole-hole pairing (though rather short-range along the chain) with an "approximate" $d_{x^2-y^2}$ symmetry [24]. These materials, being lightly doped insulators, show also features of a large metal-like Fermi surface. Since the 1D SL phase precedes the superconducting transition,
then it is quite reasonable to consider similar mechanisms being
responsible for superconductivity in both underdoped cuprates and in SL
with even $n_c$ [24]. The short-range correlation for the pairs along the chains in even-leg SLs [25]  contradicts ARPES and corner-junction SQUID measurements carried out at $T<T_c(p)$. This can be explained by a substantial modification of these correlations due to the onset at $T \approx T_c(p)$ of the Josephson-like coupling between the stripes [3], resulting eventually in an effective recovery of the 2D-character of the Cu-0 system at $T<T_c(p)$. The Josephson-like coupling between the 1D SL
results in the onset of superconductivity at $T_c(p)$ with the $T_c(p)$
value increasing with the hole doping. Isolated 1D even-chain SLs, most probably, have no chance to develop along the chain a real macroscopic superconducting coherence of bosons, formed along rungs in the SL. From this point of view, recent pulsed field data [26] can be interpreted as an experimental evidence of the insulating ground state of field-decoupled SLs in underdoped cuprates.

The 1D SL phase becomes unstable at high temperatures, when entropy effects cause stripe meandering and eventually
destroy the 1D stripes [16], and at high doping levels, when the distance
between the stripes becomes so small that the Mott insulator phase between
stripes collapses, thus leading to a recovery of the 2D regime.  As a
result, in optimally doped and overdoped regime the superconducting
transition takes place when the CuO$_2$ layer is in the regime of the 2D
doped Heisenberg system. Therefore, the 1D SL phase seems to exist only in underdoped cuprates in the temperature window between $T^*(p)$ and $T_c(p)$.

I would like to thank J.~Vanacken, L.~Trappeniers, V.~Bruyndoncx and R.~Provoost for their
technical help in preparing this paper and Y.~Bruynseraede for valuable
discussions.  This work was supported by the GOA and FWO-Vlaanderen
programs.
\newpage
\begin{enumerate}
\item B.~Batlogg, Physica C {\bf 282-287}, XXIV (1997).
\item A.V.~Chubukov, D.~Pines, and B.P.~Stojkovic, J.\ Phys.: Condens.\
Matter {\bf 8}, 10017 (1996).
\item V.J.~Emery, S.A.~Kivelson, and O.~Zachar, Physica C {\bf 282-287},
174 (1997).
\item J.L.~Tallon, G.V.M.~Williams, N.E.~Flower and C.~Bernhard, Physica C
{\bf 282-287}, 236 (1997).
\item S.~Maekawa and T.~Tohyama, Physica C {\bf 282-287}, 286 (1997).
\item H.Y.~Hwang, B.~Batlogg, H.~Takagi, H.L.~Kao, J.~Kwo, R.J.~Cava,
J.J.~Krajewski, and W.F.~Pock, Jr., Phys.\ Rev.\ Lett.\ {\bf 72}, 2636
(1994).
\item B.~Wuyts, V.V.~Moshchalkov, and Y.~Bruynseraede, Phys.\ Rev.\ B {\bf
53}, 9418 (1996).
\item T.~Ito, K.~Takenaka, and S.~Uchida, Phys.\ Rev.\ Lett.\ {\bf 70},
3995 (1993).
\item V.~Barzykin and D.~Pines, Phys.\ Rev.\ B {\bf 52}, 13585 (1995).
\item H.~Yasuoka, Physica C {\bf 282-287}, 119 (1997).
\item V.V.~Moshchalkov, Sol.\ St.\ Comm.\ {\bf 86}, 715 (1993).
\item A.A.~Abrikosov: ``Fundamentals of the Theory of Metals'',
North-Holland, Amsterdam, 1988.
\item S.~Chakravarty, B.I.~Halperin, and D.R.~Nelson, Phys.\ Rev.\ B {\bf
39}, 2344 (1989).
\item P.~Hasenfratz and F.~Niedermayer, Physics Lett.\ B {\bf 268}, 231
(1991).
\item M.~Greven, R.J.~Birgeneau, and U.-J.~Wiese, Phys.\ Rev.\ Lett.\ {\bf
77}, 1865 (1996).
\item J.~Zaanen and W.~van~Saarloos, Physica C {\bf 282-287}, 178 (1997).
\item J.M.~Tranquada, Physica C {\bf 282-287}, 166 (1997);\newline
J.M.~Tranquada et al., Nature {\bf 375}, 561 (1995);\newline
Phys.\ Rev.\ Lett.\ {\bf 73}, 338 (1997).
\item T.~Nagata et al., Physica C {\bf 282-287}, 153 (1995).
\item M.~Takano et al., Physica C {\bf 282-287}, 149 (1995).
\item J.R.~Schrieffer, S.C.~Zhang, and X.G.~Wen, Phys.\ Rev.\ Lett.\ {\bf
60}, 944 (1988).
\item H.~Ding et al., Phys.\ Rev.\ Lett.\ {\bf 76}, 1533 (1996).
\item P.W.~Anderson, Science {\bf 235}, 1196 (1987).
\item H.~Fukuyama and H.~Kohno, Physica C {\bf 282-287}, 124 (1997).
\item E.~Dagotto and T.M.~Rice, Science {\bf 271}, 618 (1996).
\item S.~Chakravarty, Science {\bf 278}, 1412 (1997).
\item Y.~Ando et al., Phys.\ Rev.\ Lett.\ {\bf 77}, 2065 (1996).
\end{enumerate}
\newpage
\begin{center}
\section*{Figure Captions}
\end{center}

\parbox[t]{1.5cm}{
Fig.~1}\ \ \parbox[t]{13.5cm}{ Temperature dependence of resistivity for
the even-chain
spin-ladder single crystal Sr$_{2.5}$Ca$_{11.5}$Cu$_{24}$O$_{41}$:
circles~-- experimental data of Nagata et al.\ [18], solid line~--
fit using Eq.~\ref{5} describing resistivity of the 1D even-chain
spin-ladders.}
\vspace*{5mm}

\parbox[t]{1.5cm}{
Fig.~2} \ \ \parbox[t]{13.5cm}{(a)~Scaled in-plane resistivity for the
YBa$_2$Cu$_3$O$_x$ system (after Ref.~7).\newline
(b)~Circles~-- the averaged data points of the scaled in-plane resistivity,
solid lines~-- the fit using Eq.~\ref{5} at $T \lesssim T^*$ (resistivity
of the 1D even-chain spin-ladder at $T \lesssim T^*$) and Eq.~\ref{1} and
Eq.~\ref{3} at $T \gtrsim T^*$ (resistivity of the 2D Heisenberg systems).}
\vspace*{5mm}

\parbox[t]{1.5cm}{
Fig.~3} \ \ \parbox[t]{13.5cm}{Scaled Knight shift data (open circles) of
Alloul (Phys.\ Rev.\ Lett.\ {\bf
63}, 689 (1989)), after Ref.~7.  The scaling parameter $T_0$ is the same
as the one used for scaling of in-plane resistivity of YBa$_2$Cu$_3$O$_x$
samples (Fig.~2a).  The solid line is the fit with Eq.~\ref{6}.}
\vspace*{5mm}

\parbox[t]{1.5cm}{
Fig.~4} \ \ \parbox[t]{13.5cm}{Schematic phase diagram of layered high
$T_c$ cuprates.}
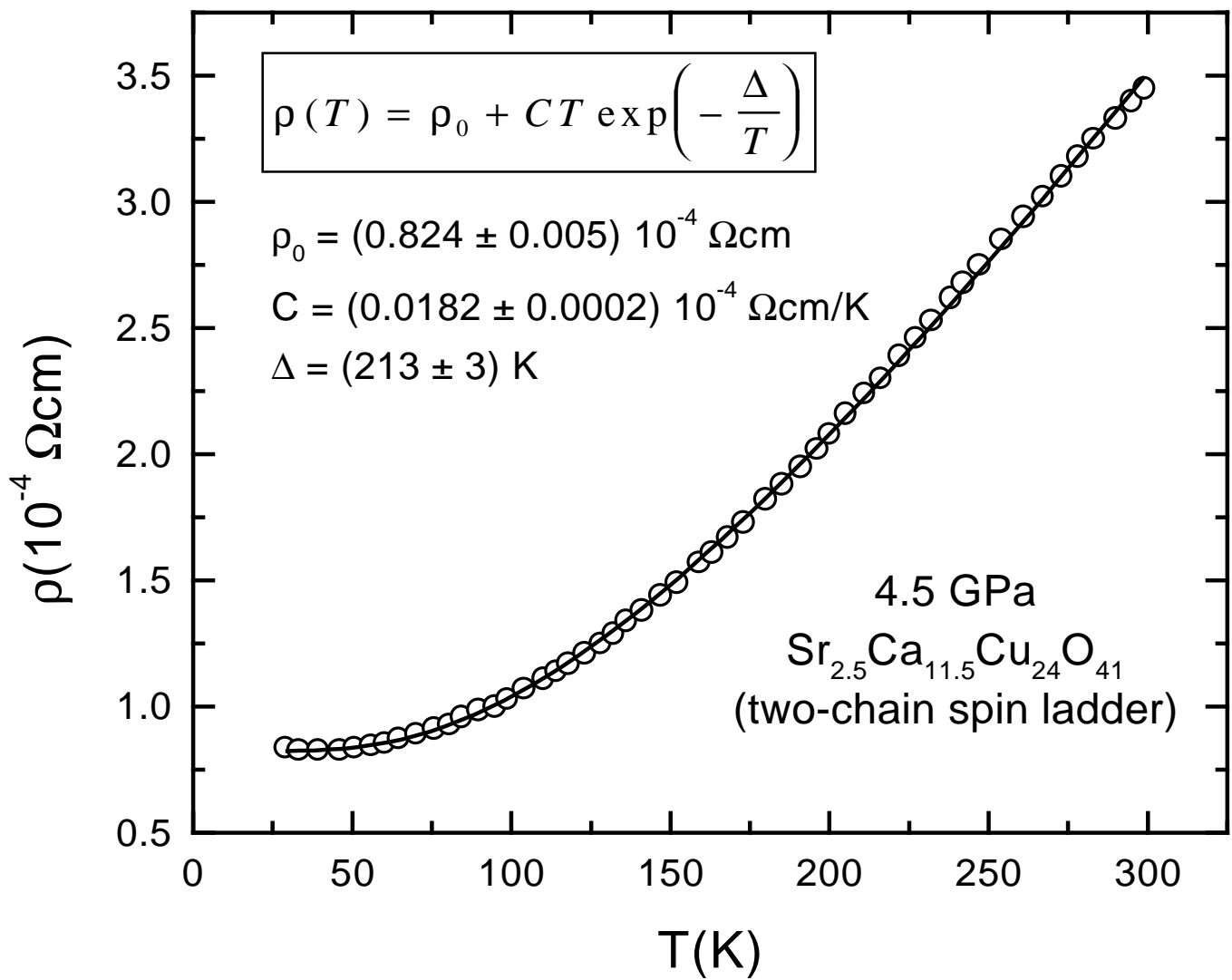
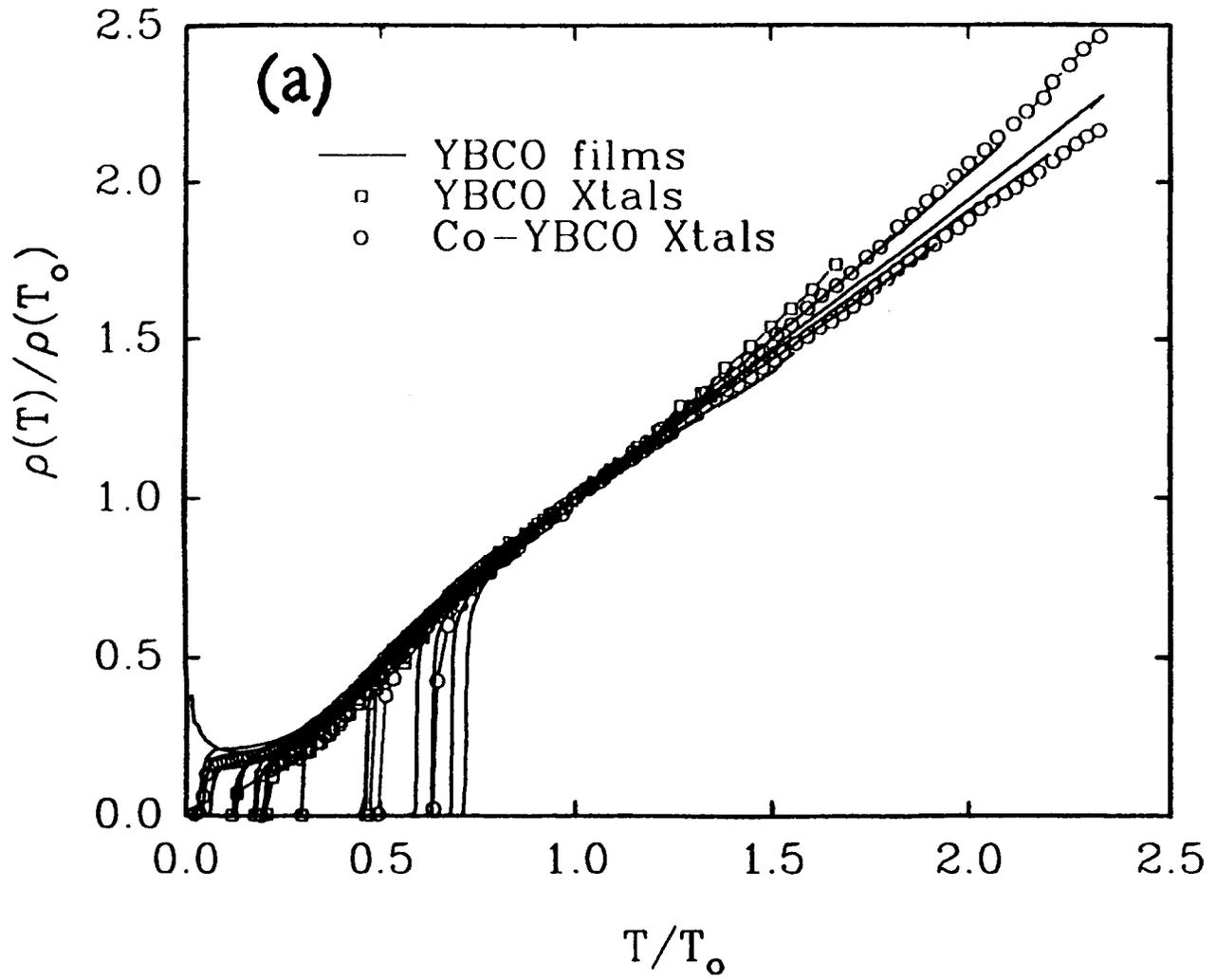
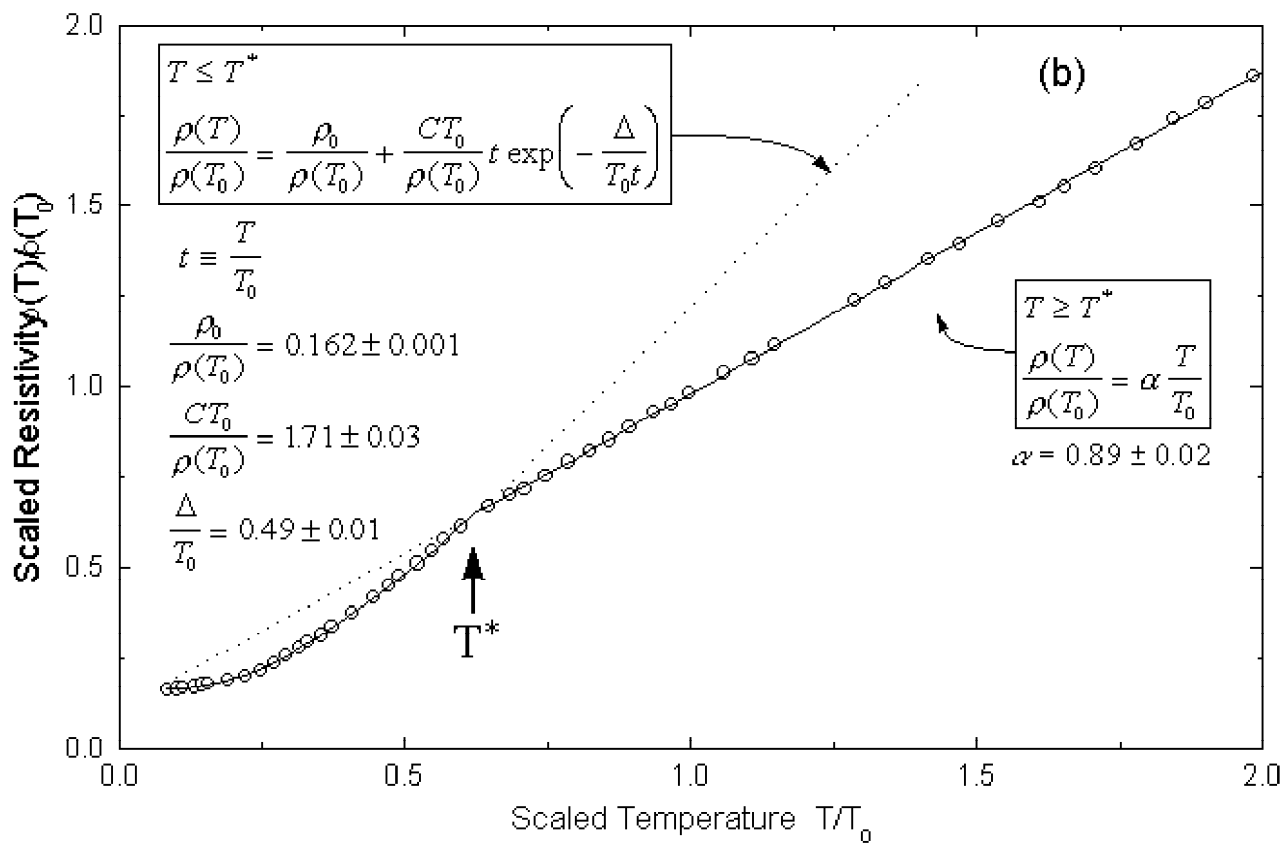
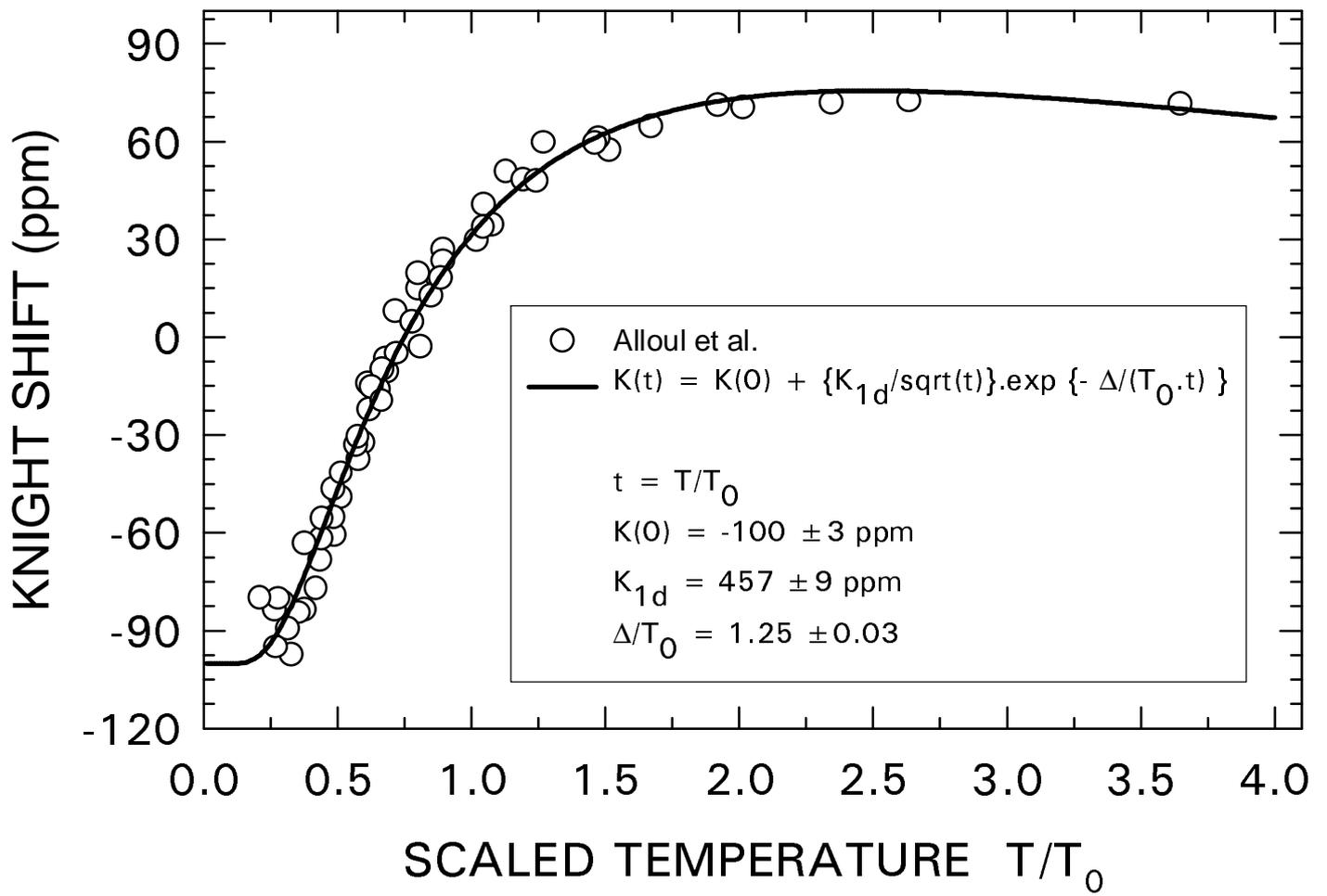
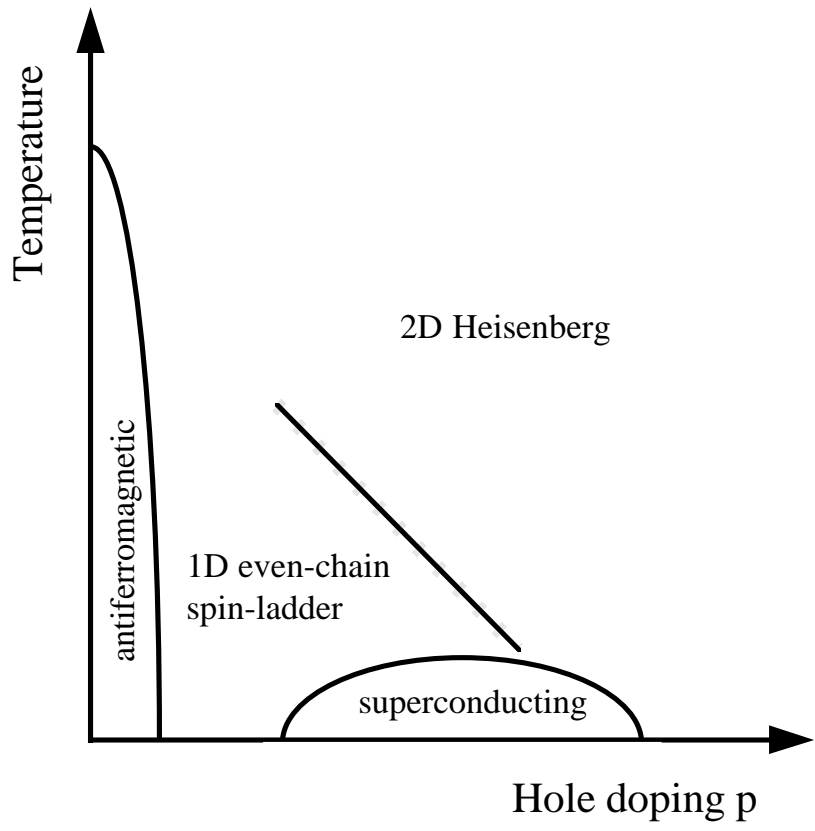
\end{document}